%% file: main.tex
\documentclass[11pt]{article}
\usepackage{fullpage}
\usepackage{color}
\usepackage{multirow}
\usepackage{float}
\usepackage{graphicx}
\usepackage{caption}
\usepackage{subcaption}
\usepackage{booktabs}
\usepackage{amsmath}
\usepackage{amsfonts}
\usepackage{algpseudocode}
\usepackage{algorithm, algorithmicx}
\usepackage{hyperref}
\usepackage{enumitem}

\usepackage{graphics} 

\title{
  Matrix Factorizations at Scale: a Comparison of Scientific Data Analytics in Spark and C+MPI Using Three Case Studies 
}

\author{Alex Gittens\thanks{ICSI and Department of Statistics, UC Berkeley} \and Aditya Devarakonda\thanks{EECS, UC Berkeley} \and Evan Racah\thanks{NERSC, Lawrence Berkeley National Laboratory} \and Michael Ringenburg\thanks{Cray, Inc.} \and  Lisa Gerhardt\footnotemark[3] \and Jey Kottalam\footnotemark[2] \and Jialin Liu\footnotemark[3] \and Kristyn Maschhoff$^{4}$ \and Shane Canon\footnotemark[3] \and Jatin Chhugani\thanks{Hiperform Consulting LLC} \and Pramod Sharma\footnotemark[4] \and Jiyan Yang\thanks{ICME, Stanford University} \and James Demmel\thanks{EECS and Math, UC Berkeley} \and Jim Harrell\footnotemark[4] \and Venkat Krishnamurthy\footnotemark[4] \and Michael W. Mahoney\footnotemark[1] \and Prabhat\footnotemark[3]
}

\date{August 23, 2016}

\begin{document}
\maketitle

\input{text/abstract}

\section{Introduction}
\label{sec:introduction}
\input{text/intro}

\section{Science Drivers and Data sets}
\label{sec:science}
\input{text/datasets}

\section{Methods}
\label{sec:methods}
\input{text/methods}

\section{Implementation}
\label{sec:implementation}
\input{text/implementation}

\section{Experimental Setup}
\label{sec:setup}
\input{text/expsetup}

\section{Results}
\label{sec:results}
\input{text/results}

\section{Lessons Learned}
\label{sec:lessons}
\input{text/sparklessons}

\section{Conclusion}
\label{sec:conclusions}
\input{text/conclusion}

\section*{Acknowledgments}
\input{text/ack}

\nocite{*}
\bibliographystyle{abbrv}
\bibliography{refs}

\end{document}

%% file: text/abstract.tex
\begin{abstract}
We explore the trade-offs of performing linear algebra using Apache Spark, compared to traditional C and MPI implementations on HPC platforms. Spark is designed for data analytics on cluster computing platforms with access to local disks and is optimized for data-parallel tasks. We examine three widely-used and important matrix factorizations: NMF (for physical plausability), PCA (for its ubiquity) and CX (for data interpretability). We apply these methods to TB-sized problems in particle physics, climate modeling and bioimaging.  The data matrices are tall-and-skinny which enable the algorithms to map conveniently into Spark's data-parallel model. We perform scaling experiments on up to 1600 Cray XC40 nodes, describe the sources of slowdowns, and provide tuning guidance to obtain high performance. 
\end{abstract}

%% file: text/intro.tex
Modern scientific progress relies upon experimental devices, observational instruments, and scientific simulations. These important modalities produce massive amounts of complex data: in High Energy Physics, the LHC project produces PBs of data; smaller-scale projects such as Daya Bay produce 100s of TBs. In Climate science, the worldwide community relies upon distributed access to the CMIP-5 archive, which is several PBs in size. In  Biosciences, multi-modal imagers can acquire 100GBs-TBs of data. These projects spend a considerable amount of effort in data movement and data management issues, but the key step in gaining scientific insights is \emph{data analytics}. Several scientific domains are currently rate-limited by access to productive and performant data analytics tools. 

Some of the most important classes of scientific data analytics methods rely on matrix algorithms. 
Matrices provide a convenient mathematical structure with which to model data arising in a broad range of applications: an $m \times n$ real-valued matrix $A$ provides a natural structure for encoding information about $m$ objects, each of which is described by $n$ features; alternatively, an $n \times n$ real-values matrix $A$ can be used to describe the correlations between all pairs of $n$ data points. 
Matrix factorizations are common in numerical analysis and scientific computing, where the emphasis is on running time, largely since they are used simply to enable the rapid solution of linear systems and related problems.
In statistical data analysis, however, matrix factorizations are typically used to obtain some form of lower-rank (and therefore simplified) approximation to the data matrix $A$ to enable better understanding the structure of the data~\cite{HMH00}.
In particular, rather than simply providing a mechanism for solving another problem quickly, the actual components making up a factorization are of prime concern.
Thus, it is of interest to understand how popular factorizations interact with other aspects of the large-scale data analysis pipeline.

Along these lines, we have recently seen substantial progress in the development and adoption of Big Data software frameworks such as Hadoop/MapReduce~\cite{DG04} and Spark~\cite{SPARK_HOTC_10}. 
These frameworks have been developed for industrial applications and commodity datacenter hardware; and they provide high productivity computing interfaces for the broader data science community.  
Ideally, the scientific data analysis and high performance computing (HPC) communities would leverage the momentum behind Hadoop and Spark.
Unfortunately, the performance of such frameworks at scale on conventional HPC hardware has not been investigated extensively. 
For linear algebraic computations more broadly, and matrix factorizations in particular, there is a gap between the performance of  well-established libraries (ScaLAPACK, LAPACK, BLAS, PLASMA, MAGMA, etc.~\cite{lapack99,PlasmaMagma2009}) and toolkits available in Spark. 
Our work takes on the important task of testing nontrivial linear algebra and matrix factorization computations in Spark for real-world, large-scale scientific data analysis applications. We compare and contrast its performance with C+MPI implementations on HPC hardware. The main contributions of this paper are as follows:
\begin{itemize}
  \item{We develop parallel versions of three leading matrix factorizations (PCA, NMF, CX) in Spark and C+MPI; and we apply them to several TB-sized scientific data sets.}
\item{We conduct strong scaling tests on a XC40 system, and we test the scaling of Spark on up to 1600 nodes.}
\item{We characterize the performance gap between Spark and C+MPI for matrix factorizations, and  comment on opportunities for future work in Spark to better address large scale scientific data analytics on HPC platforms.}
\end{itemize}


%% file: text/datasets.tex
\begin{figure*}
\centering
\begin{subfigure}[b]{0.32\textwidth}
\includegraphics[width=\textwidth]{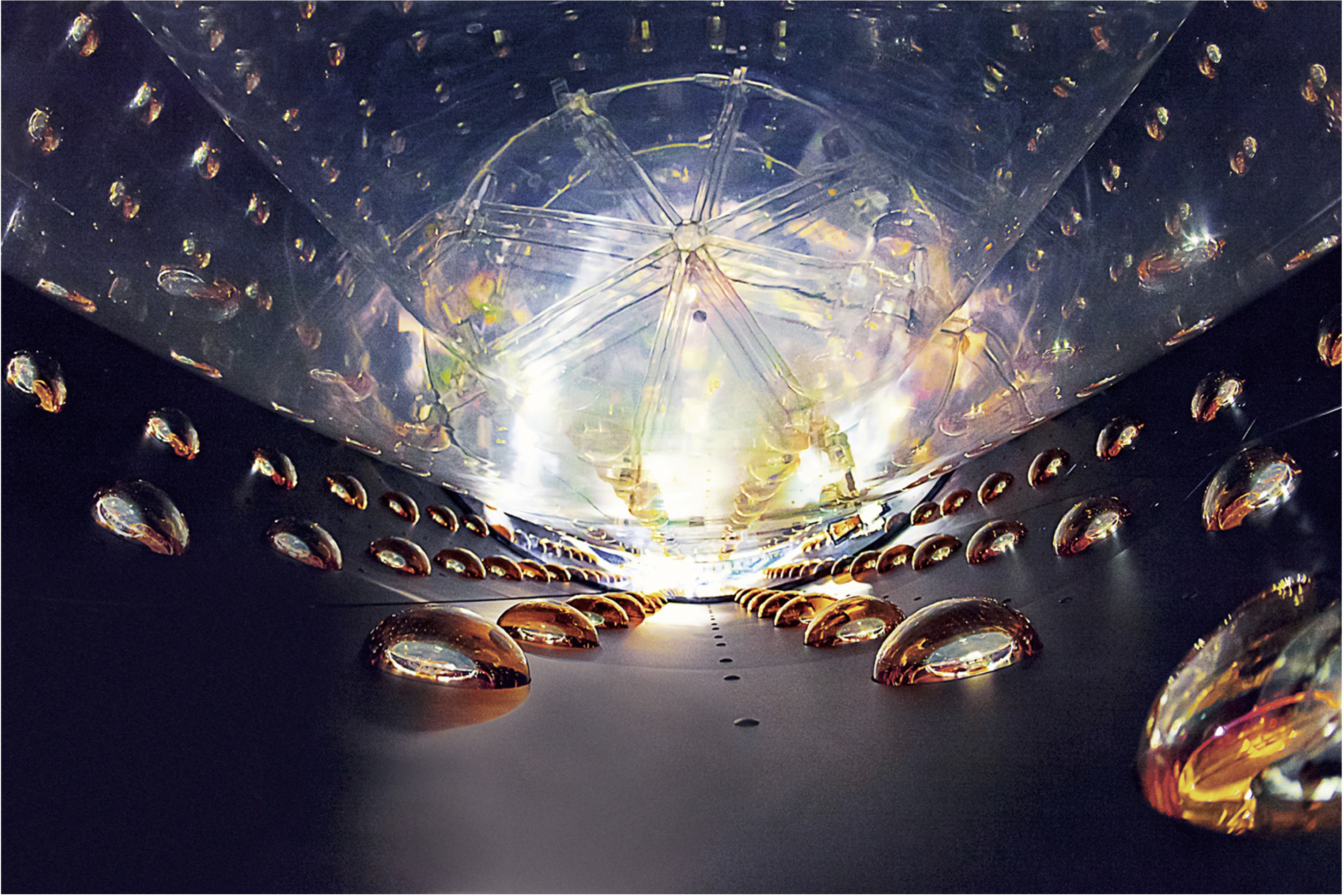}
\caption{Daya Bay Neutrino Experiment}
\label{fig:dayabay}
\end{subfigure}
\begin{subfigure}[b]{0.42\textwidth}
\includegraphics[width=\textwidth]{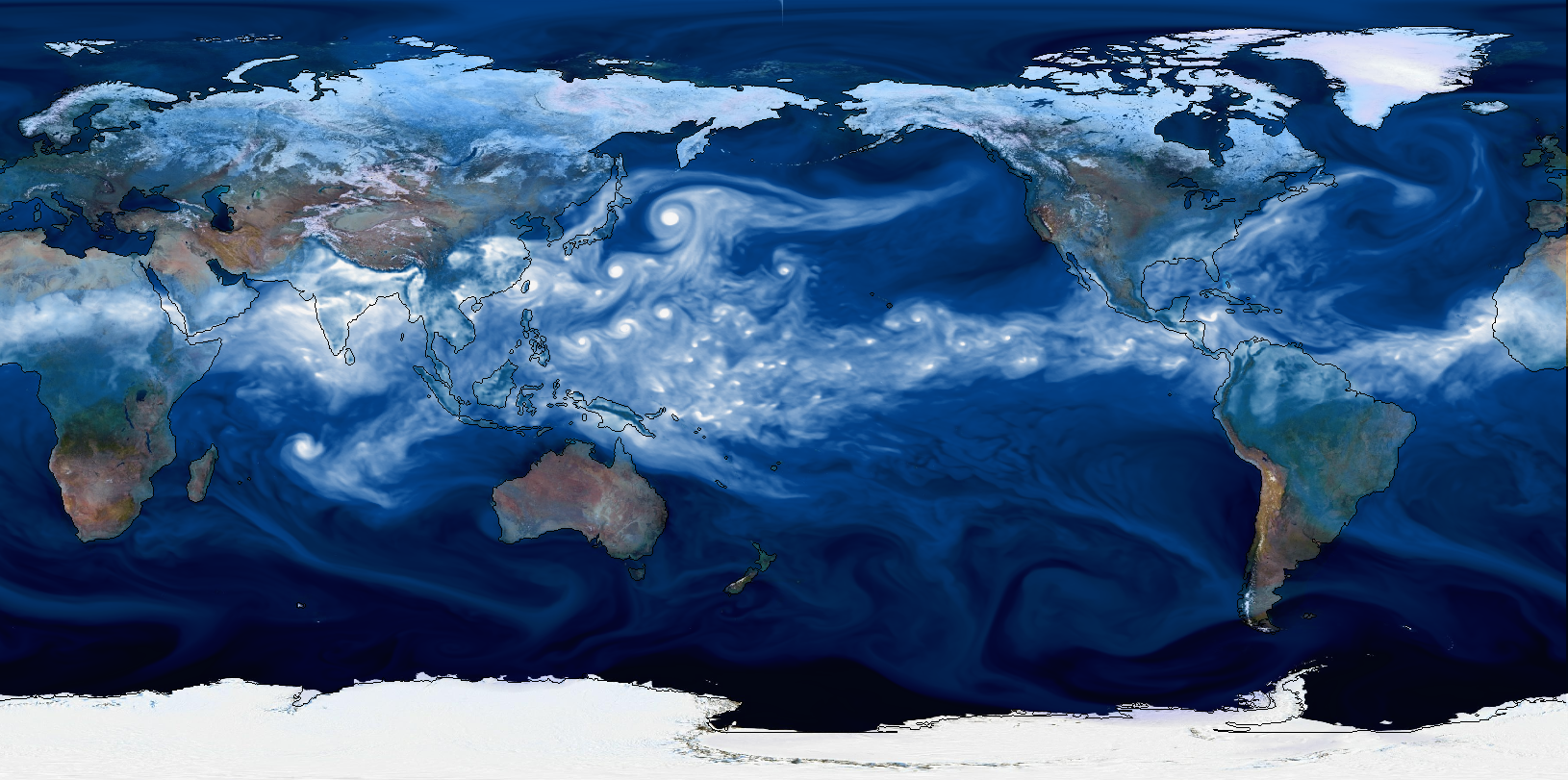}
\caption{CAM5 Simulation}
\label{fig:cam5}
\end{subfigure}
\begin{subfigure}[b]{0.22\textwidth}
\includegraphics[width=\textwidth]{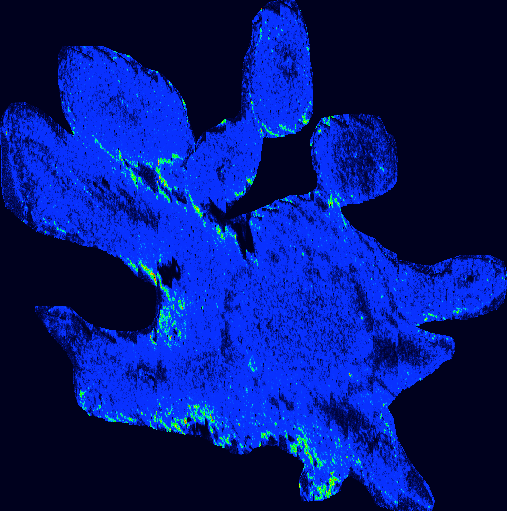}
\caption{Mass-Spec Imaging}
\label{fig:mass-spec}
\end{subfigure}
\caption{Sources of various data sets used in this study}\label{fig:datasets}
\end{figure*}
In this study, we choose leading data sets from experimental, observational, and simulation sources, and we identify associated data analytics challenges. 
The properties of these data sets are summarized in Table~\ref{table:datasets}.

\begin{table}[ht]
\centering
\caption{Summary of the matrices used in our study}
\label{table:datasets}
\begin{tabular}{p{2cm}lllll@{}}
\toprule
Science Area & Format/Files & Dimensions & Size  \\ \midrule
MSI      & Parquet/2880        &  $8,258,911 \times 131,048$          & 1.1TB  \\
Daya Bay & HDF5/1      &   $1,099,413,914 \times 192$         & 1.6TB \\
Ocean              & HDF5/1      &  $6,349,676 \times 46,715$          & 2.2TB \\
Atmosphere           & HDF5/1       & $26,542,080 \times 81,600$           & 16TB \\ \bottomrule
\end{tabular}
\end{table}

\paragraph{The Daya Bay Neutrino Experiment.}
The Daya Bay Neutrino Experiment (Figure \ref{fig:dayabay}) is situated on the southern coast of China. It detects antineutrinos produced by the Ling Ao and Daya Bay nuclear power plants and uses them to measure theta-13, a fundamental constant that helps describe the flavor oscillation of neutrinos. In 2012 the Daya Bay experiment measured this with unprecedented precision. This result was named one of the Science magazines top ten breakthroughs of 2012, and this measurement has since been refined considerably \cite{dayabay15}.

The Daya Bay Experiment consists of eight smaller detectors, each with 192 photomultiplier tubes that detect light generated by interaction of anti-neutrinos from the nearby nuclear reactors. Each detector records the total charge in each of the 192 photomultiplier tubes, as well as the time the charge was detected. For this analysis we used a data array comprised of the sum of the charge for every photomultiplier tube from each Daya Bay detector. This data is well suited to NMF analysis since accumulated charge will always be positive (with the exception of a few mis-calibrated values). The extracted data was stored as HDF5 files with 192 columns, one for each photomultiplier tube, and a different row for each discrete event in the detectors. The resulting data set is a sparse 1.6 TB matrix. The specific analytics problem that we tackle in this paper is that of finding characteristic patterns or signatures corresponding to various event types. Successfully ``segmenting'' and classifying a multiyear long timeseries into meaningful events can dramatically improve the productivity of scientists and enable them to focus on anomalies, which can in turn result in new physics results.

\paragraph{Climate Science.}
Climate scientists rely on HPC simulations to understand past, present and future climate regimes. Vast amounts of 3D data (corresponding to atmospheric and ocean processes) are readily available in the community. Traditionally, the lack of scalable analytics methods and tools has prevented the community from analyzing full 3D fields; typical analysis is thus performed only on 2D spatial averages or slices. The most widely used tool for extracting important patterns from the measurements of atmospheric and oceanic variables is the Empirical Orthogonal Function (EOF) technique. EOFs are popular because of their simplicity and their ability to reduce the dimensionality of large nonlinear, high-dimensional systems into fewer dimensions while preserving the most important patterns of variations in the measurements. Mathematically, EOFs are exactly PCA decompositions.

In this study, we consider the Climate Forecast System Reanalysis Product~\cite{saha:2010}. Global Ocean temperature data, spatially resolved at 360 x 720 x 41 (latitude x longitude x depth) and 6-hour temporal resolution is analyzed. The CFSR data set yields a dense 2.2TB matrix. We also process a CAM5 0.25-degree atmospheric humidity data set~\cite{wehner:2014} (Figure~\ref{fig:cam5}). The grid is 768 x 1158 x 30 (latitude x longitude x height) and data is stored every 3 hours. The CAM5 data set spans 28 years, and it yields a dense 16TB matrix. The specific analytics problem that we tackle is finding the principal causes of variability in large scale 3D fields. A better understanding of the dynamics of large-scale modes of variability in the ocean and atmosphere may be extracted from full 3D EOFs.

\paragraph{Mass-Spectrometry Imaging.}
Mass spectrometry measures ions derived from the molecules present in a biological sample. Spectra of the ions are acquired at each location (pixel) of a sample, allowing for the collection of spatially resolved mass spectra. This mode of analysis is known as \textit{mass spectrometry imaging (MSI)}. The addition of \textit{ion-mobility separation (IMS)} to MSI adds another dimension, drift time.  The combination of IMS with MSI is finding increasing applications in the study of disease diagnostics, plant engineering, and microbial interactions. Unfortunately, the scale of MSI data and complexity of analysis presents a significant challenge to scientists: a single 2D-image may be many gigabytes and comparison of multiple images is beyond the processing capabilities available to many scientists. The addition of IMS exacerbates these problems. 

We analyze one of the largest (1TB sized) mass-spec imaging data sets in the field, obtained from a sample of the plant {\it Lewis Dalisay Peltatum} (Figure \ref{fig:mass-spec}). The MSI measurements are formed into a sparse matrix whose rows and columns correspond to pixel and ($\tau$, $m/z$) values of ions, respectively. Here $\tau$ denotes drift time and $m/z$ is the mass-to-charge ratio. The sheer size of this data set has previously made complex analytics intractable. CX decompositions allow for the possibility of identifying small numbers of columns (ions) in the original data that reliably explain a large portion of the variation in the data.

%% file: text/methods.tex
Given an $m \times n$ data matrix $A$, low-rank matrix factorization methods aim to find two or more smaller matrices such that their product is a good approximation to $A$.
That is, they aim to find matrices $Y$ and $Z$ such that
\begin{equation*}
 \label{eqn:apprx}
    \underset{m\times n}{A} \approx \underset{m\times k}{Y} \times \underset{k\times n}{Z}. 
\end{equation*}
Low-rank matrix factorization methods are important tools in linear algebra and numerical analysis, and they find use in a variety of scientific fields and scientific computing. These methods have the following advantages:
\begin{itemize}
  \item They are often useful in data compression, as smaller matrices can be stored more efficiently.
  \item In some cases, the results of analysis using them are more interpretable. For example, in imaging analysis, the original images can be reconstructed using linear combination of basis images.
  \item They can be viewed as a basic dimension reduction technique.
  \item In many modern applications, data sets containing a massive number of rows or columns are becoming more common, which makes it difficult for data visualization or applying classic algorithms, but low-rank approximation methods express every data point in a low-dimensional space defined by only a few features.
\end{itemize}
Throughout, we assume the data matrix $A$ has size $m \times n$ and rank $r$, with $r \ll n \ll m$; this ``tall-skinny'', highly rectangular setting is common in practice. 

Matrix factorizations are also widely-used in statistical data analysis~\cite{HMH00}.
Depending on the particular application, various low-rank factorization techniques are of interest. Popular choices include the singular value decomposition~\cite{GVL96}, principal component analysis~\cite{pcaBook}, rank-revealing QR factorization~\cite{GE96}, nonnegative matrix factorization (NMF) ~\cite{NMFalg}, and CX/CUR decompositions~\cite{CUR_PNAS}. In this work, we consider the PCA decomposition, due to its ubiquity, as well as the NMF and CX/CUR decompositions, due to their usefulness in scalable and interpretable data analysis. In the remainder of the section, we briefly describe these decompositions and the algorithms we used in our implementations, and we also discuss related implementations. 

\paragraph{Prior Work.}The body of theoretical and practical work surrounding distributed low-rank matrix factorization is large and continuously growing. The HPC community has produced many high quality packages specifically for computing partial SVDs of large matrices: \textsc{Propack}~\cite{larsen1998lanczos}, \textsc{Blopex}~\cite{blopex2007}, and \textsc{Anasazi}~\cite{baker2009anasazi}, among others. We refer the interested reader to~\cite{hernandez2009survey} for a well-written survey. As far as we are aware, there are no published HPC codes for computing CX decompositions, but several HPC codes exist for NMF factorization~\cite{kannan2016high}. 

The machine learning community has produced many packages for computing a variety of low-rank decompositions, including NMF and PCA, typically using either an alternating least squares (ALS) or a stochastic gradient descent approach~\cite{gemulla2011large,yun2014nomad,koren2009matrix}. ALS algorithms can produce high precision decompositions, but have a high computational and communication cost, while SGD algorithms produce low precision decompositions with comparatively lower costs. We mention a few of the high-visibility efforts in this space. The earlier work~\cite{liu2010distributed} developed and studied a distributed implementation of the NMF for general matrices under the Hadoop framework, while~\cite{benson2014scalable} introduced a scalable NMF algorithm that is particularly efficient when applied to tall-and-skinny matrices. We implemented a variant of the latter algorithm in Spark, as our data matrices are tall-and-skinny. The widely used MLLib library, packaged with Spark itself, provides some linear algebra datatypes (vectors and matrices) and implementations of basic linear algebra routines~\cite{meng2016mllib}; we note that the PCA algorithm implemented in MLLib is almost identical to our concurrently developed implementation. The Sparkler system introduces a memory abstraction to the Spark framework which allows for increased efficiency in computing low-rank factorizations via distributed SGD, ~\cite{Li2013sparkler}, but such factorizations are not appropriate for scientific applications which require high precision.

The 2011 report on the DOE Magellan cloud computing project~\cite{magellan2011} discusses qualitative experience implementing numerical linear algebra in Hadoop, specifically relating to the tall-skinny QR algorithm. Our contribution is the provision of, for the first time, a detailed investigation of the scalability of three low-rank factorizations using the linear algebra tools and bindings provided in Spark's baseline MLLib~\cite{meng2016mllib} and MLMatrix~\cite{zadeh2016matrix} libraries. By identifying the causes of the slow-downs in these algorithms that exhibit different bottlenecks (e.g.~I/O time versus synchronization overheads), we provide a clear indication of the issues that one encounters attempting to do serious distributed linear algebra using Spark.
To ensure that our comparison of Spark to MPI is fair, we implemented the same algorithms in Spark and MPI, drawing on a common set of numerical linear algebra libaries for which Spark bindings are readily available (\textsc{Blas}, \textsc{Lapack}, and \textsc{Arpack}).

\paragraph{Principal Components Analysis.} 
Principal component analysis (PCA) is closely related to the singular value decomposition (SVD).
In particular, the PCA decomposition of a matrix $A$ is the SVD of the matrix formed by centering each column of $A$ (i.e., removing the mean of each column) and considering $A^TA$ (or $AA^T$).
The SVD is the most fundamental low-rank matrix factorization because it provides the best low-rank matrix approximation with respect to any unitarily invariant matrix norm.
In particular, for any target rank $k \leq r$, the SVD provides the minimizer of the optimization problem
\begin{equation}
 \label{eqn:obj}
  \min_{\text{rank}(\tilde A) = k} \| A - \tilde A \|_F,
\end{equation}
where the Frobenius norm $\| \cdot \|_F$ is defined as $\|X\|_F^2 =
\sum_{i=1}^m \sum_{j=1}^n X_{ij}^2 $.
The solution
to~\eqref{eqn:obj} is given by the truncated SVD, i.e., $A_k = U_k \Sigma_k
V_k^T$, where the columns of $U_k$ and $V_k$ are the top $k$ {\it left and right singular vectors}, respectively, and $\Sigma_k$ is a 
diagonal matrix containing the corresponding top $k$ {\it singular values}.


\begin{algorithm}[tb]
    \caption{\textsc{PCA} Algorithm}
    \label{alg:pca}
    \begin{algorithmic}[1]
      \Require $A \in \mathbb{R}^{m\times n}$, rank parameter $k \leq \textrm{rank}(A).$
      \Ensure $U_k \Sigma_k V_k^T = \textsc{PCA}(A, k).$
      \State Let $(V_k, \_) = \textsc{IRAM}(\textsc{MultiplyGramian}(A, \cdot), k).$
      \State Let $Y = \textsc{Multiply}(A, V_k).$
      \State Compute $(U_k, \Sigma_k, \_) = \textsc{SVD}(Y).$
    \end{algorithmic}
  \end{algorithm}
  
Direct algorithms for computing the PCA decomposition scale as $\mathcal{O}(mn^2)$, so are not feasible for the scale of the problems we consider. Instead, we use the iterative algorithm presented in Algorithm~\ref{alg:pca}: in step 1, a series of distributed matrix-vector products against $A^T A$  (\textsc{MultiplyGramian}) are used to extract $V_k$ by applying the implicitly restarted Arnoldi method (\textsc{IRAM})~\cite{lehoucq1996deflation}, then in step 2 a distributed matrix-matrix product followed by a collect is used to bring $AV_k$ to the driver. Step 3 occurs on the driver, and computes a final SVD on $AV_k$ to extract the top left singular vectors $U_k$ and the corresponding eigenvalues $\Sigma_k.$ Here QR and SVD compute the ``thin'' versions of the QR and SVD decompositions~\cite{GVL96}. (Algorithm~\ref{alg:pca} calls \textsc{MultiplyGramian}, which is summarized in Algorithm~\ref{alg:gram}).

  \begin{algorithm}[tb]
    \caption{{\sc MultiplyGramian} Algorithm}
    \label{alg:gram}
    \begin{algorithmic}[1]
      \Require $A \in \mathbb{R}^{m\times n}$, $B \in \mathbb{R}^{n\times k}$.
      \Ensure $X = A^T A B$.
      \State Initialize $X = 0$.
      \For{each row $a$ in $A$}
          \State $X \gets X + a a^T B$.
      \EndFor
    \end{algorithmic}
\end{algorithm}

\textsc{ML-Lib}, Spark's machine learning library, provides implementations of the SVD and PCA, as well as an alternating least squares algorithm for low-rank factorization~\cite{meng2015mllib} (the PCA algorithm used in ML-Lib is very similar to Algorithm~\ref{alg:pca}, but explicitly computes $A^TA$). Similarly, the Apache Mahout project provides Hadoop and Spark implementations of the PCA, \textsc{RandomizedSVD}, and ALS algorithms. However, to our knowledge, there are no published investigations into the impact Spark or MapReduce's infrastructure has on the performance of these algorithms.
  
\paragraph{Nonnegative Matrix Factorization.}
Although the PCA provides a mathematically optimal low-rank decomposition in the sense of~\eqref{eqn:obj}, in some scientific applications retaining sparseness and interpretability is as important as explaining variability. 
Various nonnegative matrix factorizations (NMFs) provide interpretable low-rank matrix decompositions when the columns of $A$ are nonnegative and can be viewed as additive superpositions of a small number of positive factors~\cite{gillis2014and}. NMF has found applications, among other places, in medical imaging~\cite{lee2001nmf}, facial recognition~\cite{guillamet2002non}, chemometrics~\cite{Paatero199723}, hyperspectral imaging~\cite{gillis2015hierarchical}, and astronomy~\cite{pauca2006nonnegative}.

\begin{algorithm}[tb]
    \caption{\textsc{NMF} Algorithm}
    \label{alg:nmf}
    \begin{algorithmic}[1]
      \Require $A \in \mathbb{R}^{m\times n}$ with $A \geq 0$, rank parameter $k \leq \textrm{rank}(A).$
      \Ensure $W H \approx A$ with $W,H \geq 0$
      \State Let $(\_, R) = \textsc{TSQR}(A).$
      \State Let $(\mathcal{K}, H) = \textsc{Xray}(R, k).$
      \State Let $W = A(:, \mathcal{K}).$
    \end{algorithmic}
  \end{algorithm}
  
The basic optimization problem solved by NMF is
\begin{equation}
\min_{W,H \geq 0} \|A - WH\|_F,
\end{equation}
where $W \in \mathbb{R}^{m \times k}$ and $H \in \mathbb{R}^{k \times n}$ are entry-wise nonnegative matrices. Typical approaches attempt to solve this non-convex problem by using block coordinate optimizations that require multiple passes over $A$~\cite{kim2014algorithms}. We adopt the one-pass algorithm of~\cite{benson2014scalable}. This approach makes the assumption that $W$ can be formed by {\it selecting} columns from $A$. In this setting, the columns of $A$ constituting $W$ as well as the corresponding $H$ can be computed directly from the (much smaller) $R$ factor in a thin QR factorization of $A$. More details are given in Algorithm~\ref{alg:nmf}: in step 1, a one pass distributed tall-skinny QR (\textsc{TSQR}) factorization~\cite{demmel12} is used to compute the $R$ factor of $A$; in step 2, which occurs on the driver, the \textsc{Xray} algorithm of~\cite{kumar13} is applied to $R$ to simultaneously compute $H$ and the column indices $\mathcal{K}$ of $W$ in $A$. Finally, $W$ can be explicitly computed once $\mathcal{K}$ is known.

The \textsc{ML-Lib} and Mahout libraries provide alternating least squares-based NMF implementations in Spark and MapReduce, respectively, and several other NMF implementations are available for the MapReduce framework~\cite{liu2010distributed,Liao201448,benson2014scalable}. We note that~\cite{benson2014scalable} introduced Algorithm~\ref{alg:nmf}. None of these works quantified the performance costs of implementing these algorithms in the Spark or MapReduce frameworks.

\paragraph{CX/CUR decompositions.}
CX (and related CUR) decompositions are
low-rank matrix decompositions that are expressed in terms of a small number of actual columns/rows, i.e, actual data elements, as opposed to eigencolumns/eigenrows.  
As such, they have been used in scientific applications where coupling analytical techniques with domain knowledge is at a premium, including
genetics~\cite{Paschou07b}, astronomy~\cite{Yip14-AJ}, and mass spectrometry imaging~\cite{YRPMB15}. To find a CX decomposition, we seek matrices $C$ and $X$ such that the approximation error $\|A-CX\|_F$ is small and $C$ is an $m\times k$ matrix comprising of $k$
actual columns of $A$ and $X$ is a $k \times n$ matrix.


The randomized algorithm of~\cite{DMM08} generates a $C$ whose approximation error is, with high probability, within a multiplicative factor of $(1+\varepsilon)$ of the optimal error obtainable with a low-rank decomposition:
\[
\|A - CX\|_F \leq (1+ \varepsilon) \|A - A_k\|_F.
\]
This algorithm takes as input the (approximate or exact) \emph{leverage scores} of the columns of $A.$ The leverage score of the $j$-th column of $A$ is defined in terms of $V_k$, the matrix of top k right singular vectors:
  \begin{equation}
    \label{eqn:lev}
     \ell_i = \sum_{j=1}^k (V_k) _{ij}^2;
   \end{equation}
the leverage scores can be approximated using an approximation to $V_k.$ The CX algorithm uses those scores as a sampling distribution to select $k$ columns from $A$ to form $C$; once the matrix $C$ is determined, the optimal matrix $X$ that minimizes $\|A-CX\|_F$ can be computed accordingly; see~\cite{DMM08} for the details of this construction.


The computational cost of the CX decomposition is determined by the cost of computing $V_k$ exactly or approximately. To approximate $V_k$, we use the \textsc{RandomizedSVD} algorithm introduced in \cite{MRT06,MRT11}. We refer the readers to \cite{HMT09_SIREV,Mah-mat-rev_BOOK} for more details. Importantly, the algorithm runs in $\mathcal{O}(mn \log k)$ time and needs only a constant number of passes over the data matrix ($q$+1), where $q$ is an input in Algorithm~\ref{alg:cx}).  The \textsc{RandomizedSVD} algorithm comprises the first nine steps of Algorithm~\ref{alg:cx}. The running time cost for \textsc{RandomizedSVD} is dominated by a distributed matrix-matrix multiplication appearing in Steps 3 and 7 of Algorithm~\ref{alg:cx}. After Step 7, $Y$ is collected the remaining computations are carried out on the driver.

\begin{algorithm}[tb]
   \caption{{\sc CX} Algorithm}
    \label{alg:cx}
    \begin{algorithmic}[1]
      \Require $A \in \mathbb{R}^{m\times n}$, \
        number of power iterations $q \ge 1$, \
        target rank $k > 0$, slack $p \ge 0$, and let $\ell=k+p$.

      \Ensure $C$.

      \State Initialize $B \in \mathbb{R}^{n\times \ell}$ by sampling $B_{ij} \sim \mathcal{N}(0, 1)$.

      \For{$q$ times}
          \State $B \gets \Call{MultiplyGramian}{A, B}$
          \State $(B, \_) \gets \Call{QR}{B}$
      \EndFor

      \State Let $Q$ be the first $k$ columns of $B$.

      \State Let $Y = \Call{Multiply}{A, Q}$.

      \State Compute $(U, \Sigma, \tilde V^T) = \Call{SVD}{Y}$.

      \State Let $V = Q \tilde V$.

	  \State Let $\ell_i = \sum_{j=1}^k v_{ij}^2$ for $i = 1, \ldots, n$.
      
      \State Define $p_i = \ell_i / \sum_{j=1}^d \ell_j$ for $i=1,\ldots,n$.
      
      \State Randomly sample $k$ columns from $A$ in i.i.d. trials, using the importance sampling distribution $\{p_i\}_{i=1}^n$ .
      \end{algorithmic}
  \end{algorithm}
To the best of our knowledge, this is the first published work to investigate the performance of the CX algorithm on any large-scale distributed/parallel platform.

%% file: text/implementation.tex
Spark is a parallel computing framework, built on the JVM, that adheres to the data parallelism model. A Spark cluster is composed of a driver process and a set of executor processes. The driver schedules and manages the work, which is carried out by the executors. The basic unit of work in Spark is called a task. A single executor has several slots for running tasks (by default, each core of an executor is mapped to one task) and runs several concurrent tasks in the course of calculations. Spark's primitive datatype is the resilient distributed data set (RDD), a distributed array that is partitioned across the executors. The user-defined code that is to be run on the Spark cluster is called an application. When an application is submitted to the cluster, the driver analyses its computation graph and breaks it up into jobs.  Each job represents an action on the data set, such as counting the number of entries, returning data set entries, or saving a data set to a file. Jobs are further broken down into stages, which are collections of tasks that execute the same code in parallel on a different subset of data. Each task operates on one partition of the RDD. Communication occurs only between stages, and takes the form of a shuffle, where all nodes communicate with each other, or a collect, where all nodes send data to the driver.

Spark employs a lazy evaluation strategy for efficiency. All Spark operations that have no immediate side-effects other than returning an RDD are deferred if possible. Instead, deferrable operations simply create an entry in the program's computation graph, recording the input dependencies and capturing any closures and values needed. This approach allows Spark to defer computations as much as possible, and when the evaluation is unavoidable the entire Spark job can be examined by Spark's scheduler. This allows the Spark execution engine to batch together related operations, optimize data locality, and perform better scheduling. A major benefit of Spark over MapReduce is the use of in-memory caching and storage so that data structures may be reused rather than being recomputed. Because Spark tracks the computation graph and the dependencies required for the generation of all RDDs, it natively provides fault-tolerance: given the lineage of the RDD, any lost partitions of that RDD can be recomputed as needed.

\paragraph{Implementing Matrix Factorizations in Spark.}
All three matrix factorizations store the matrices in a row-partitioned format. This enables us to use data parallel algorithms and match Spark's data parallel model.

The \textsc{MultiplyGramian} algorithm is the computational core of the PCA and CX algorithms.
This algorithm is applied efficiently in a distributed fashion by observing that if the $i$-th executor of $\ell$ stores the block of the rows of $A$ denoted by $A_{(i)},$ then $A^TA B = \sum_{i=1}^\ell A_{(i)}^T A_{(i)} B.$ Thus \textsc{MultiplyGramian} requires only one round of communication.  The local linear algebra primitives \textsc{QR} and \textsc{SVD} needed for PCA and CX are computed using the \textsc{LAPACK} bindings of the Breeze numerical linear algebra library.  The \textsc{Netlib-Java} binding of the \text{ARPACK} library supplies the \textsc{IRAM} primitive required by the PCA algorithm. 

The NMF algorithm has as its core the tall-skinny QR factorization, which is computed using a tree reduction over the row-block partitioned $A$.
We used the \textsc{TSQR} implementation available in the \textsc{Ml-Matrix} package. To implement the \textsc{XRay} algorithm, we use the \textsc{ML-Lib} non-negative least squares solver.

\paragraph{Implementing Matrix Factorizations in C+MPI.}
NMF, PCA and CX require linear algebra kernels that are available in widely-used libraries such as Intel MKL, Cray LibSci, and arpack-ng. We use these three libraries in our implementations of the matrix factorizations. The data matrices are represented as 1D arrays of double-precision floating point numbers and are partitioned across multiple nodes using a block row partitioned layout. The 1D layout enables us to use matrix-vector products and TSQR as our main computational kernels. We use MPI collectives for inter-processor communication and perform independent I/O using the Cray HDF5 parallel I/O~library.

%% file: text/expsetup.tex
All performance tests reported in this paper were conducted on the Cori system at NERSC. Cori Phase I is a Cray XC40 system with 1632 dual-socket compute nodes. Each node consists of two 2.3GHz 16-core Haswell processors and 128GB of DRAM. The Cray Aries high-speed interconnect is configured in a ``Dragonfly' topology. We use a Lustre scratch filesystem with 27PB of storage, and over 700 GB/s peak I/O performance. 

\paragraph{Spark Configuration.}
We use the Standalone Cluster Manager to run the Spark cluster. This is a collection of scripts that start the driver process and use ssh to start the executor processes on each node. Once the executors are started, they communicate with the driver via akka-tcp. When an application is submitted to the Spark cluster a second java process is spawned by each executor that controls the computation for that application. Sometimes this second process will fail to start and the executor does not participate in the calculation. The exact cause of this is not well known. Running Spark in an encapsulated Shifter image reduces the rate of these failures, which suggests this could be due to a race condition in the code.

\paragraph{H5Spark: Loading HDF5 data natively into Spark.}
The Daya Bay and climate data sets are stored in HDF5. We utilized the H5Spark~\cite{h5spark-cug16} package to read this data in as one RDD object. H5Spark provides a parallel I/O interface that efficiently loads TBs of data into the workers' memory and constructs a single RDD. An MPI-like independent I/O is performed in H5Spark to balance the workload. H5Spark partially relies on the Lustre file system striping to achieve high I/O bandwidth. We chose a Lustre configuration optimal for each data set: we stored the Daya Bay data on 72 OSTs and the climate data sets on 140 OSTs, both with striping size of 1MB. 

\paragraph{Shifter.}
Shifter is a framework that delivers docker-like functionality to HPC \cite{shifter}. It works by extracting images from native formats (such as a Docker image) and converting them to a common format that is optimally tuned for the HPC environment. 
Shifter allows users with a complicated software stack to easily install them in the environment of their choosing. It also offers considerable performance improvements because metadata operations can be more efficiently cached compared to a parallel file system and users can customize the shared library cache (ldconfig) settings to optimize access to their analysis libraries. In particular, shared library performance, which has long been a pain point on Cray systems, is dramatically improved. For this analysis we used two separate Shifter images. A generic ``CCM'' image which only contained SSH functionality (which is otherwise absent by default on Cray compute nodes) and a full “spark” image which contained version 1.5.1 of Spark  compiled with OpenBLAS \cite{openblas} and SSH. The Spark image is available on Docker Hub \cite{dockerspark}. 

\paragraph{Spark Tuning Parameters.}
Shifter provides a user-controlled option to create a writeable temporary space that is private to each node. This has performance characteristics similar to a local disk. This is created by mounting a writeable loop-back mounted file system which is backed by the parallel file system. This feature is very useful for frameworks like Spark that assume the presence of a local disk that can be used to store node local temporary files and spills. Metadata operations and small I/O transactions can  be more efficiently cached on the compute node since, unlike the Lustre scratch file system, it doesn't have to maintain coherency of this file system with other nodes. Most importantly, as the Spark cluster size is scaled up, this approach helps avoid additional pressure on the Lustre Metadata Servers which are the least scalable components of the file system. Since Spark opens and closes files many times, using the loop-back mounted file system as a writable cache can improve performance \cite{scalingspark16}.

We followed general Spark guidelines for Spark configuration values. The driver and executor memory were both set to 100 GB, a value chosen to maximize the memory available for data caching and shuffling while still leaving a buffer to hedge against running the nodes out of memory.  Generally we found that fetching an RDD from another node was detrimental to performance, so we turned off speculation (a function that restarts tasks on other nodes if it looks like the task is taking longer than average). We also set the spark locality wait to two minutes, this ensures that the driver will wait at least two minutes before scheduling a task on a node that doesn't have the task's RDD. The total number of spark cores was chosen such that there was a one-to-one correspondence between spark cores and physical cores on each node (with the exception of the 50-node NMF run which used a factor of two more partitions because it ran into hash table size issues). We used the KryoSerializer for deserialization of data. We compiled Spark to use multi-threaded OpenBLAS for PCA.

\paragraph{C+MPI Tuning Parameters.}
The NMF algorithm uses the Tall-Skinny QR (TSQR) \cite{ballard14,demmel12} factorization implemented as part of the Communication-Avoiding Dense Matrix Computations (CANDMC) library \cite{Solomonik14} which links to Intel MKL for optimized BLAS routines using the Fortran interface and ensured that loops were auto-vectorized when possible. We explored multi-threading options with OpenMP but found that it did not significantly improve performance. Applying TSQR on the Daya Bay data set results in a $192 \times 192$ upper-triangular matrix. Due to the small size we utilized a sequential non-negative least squares solver by Lawson and Hanson \cite{lawson95} in the \textsc{XRay} algorithm. PCA requires EVD, SVD, matrix-vector products, and matrix-matrix products. We use arpack-ng \cite{Lehoucq97} for the SVD and link to single-threaded Cray LibSci for optimized BLAS routines using the C interface. All experiments were conducted using a flat-MPI configuration with one MPI process per physical core and disabled TurboBoost.

\paragraph{Spark Overheads.}
To report the overheads due to Spark's communication and synchronization costs, we group them into the following bins, illustrated in Figure~\ref{fig:overheads}:

\begin{figure}[thb!]
\begin{center}
\includegraphics[width=.5\textwidth]{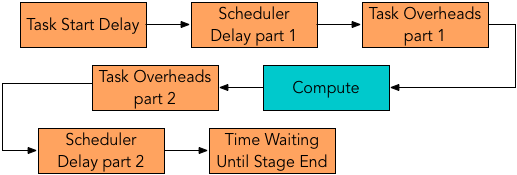}
\caption{Per task chronological breakdown of Spark overheads.}
\label{fig:overheads}
\end{center}
\end{figure}

\begin{itemize}[noitemsep,topsep=0mm]
  \item \emph{Task Start Delay}: the time between the stage start and when the driver sends the task to an executor.
  \item \emph{Scheduler Delay}: the sum of the time between when the task is sent to the executor and when it starts deserializing on the executor and the time between the completion of the serialization of the result of the task and the driver's reception of the task completion message.
  \item \emph{Task Overhead Time}: the sum of the fetch wait times, executor deserialize times, result serialization times, and shuffle write times.
  \item \emph{Time Waiting Until Stage End}: the time spent waiting on the final task in the stage to end.
\end{itemize}

%% file: text/results.tex
\subsection{NMF applied to the Daya Bay matrix}
The separable NMF algorithm we implemented fits nicely into a data parallel programming model. After the initial distributed TSQR the remainder of the algorithm is computed serially on the driver. The Daya Bay matrix is especially amenable to this approach, as the extreme aspect ratio of the data set implies that the TSQR is particularly efficient.

\paragraph{C+MPI vs. Spark.}

\begin{figure*}[thb!]
\begin{center}
\includegraphics[width=.9\textwidth]{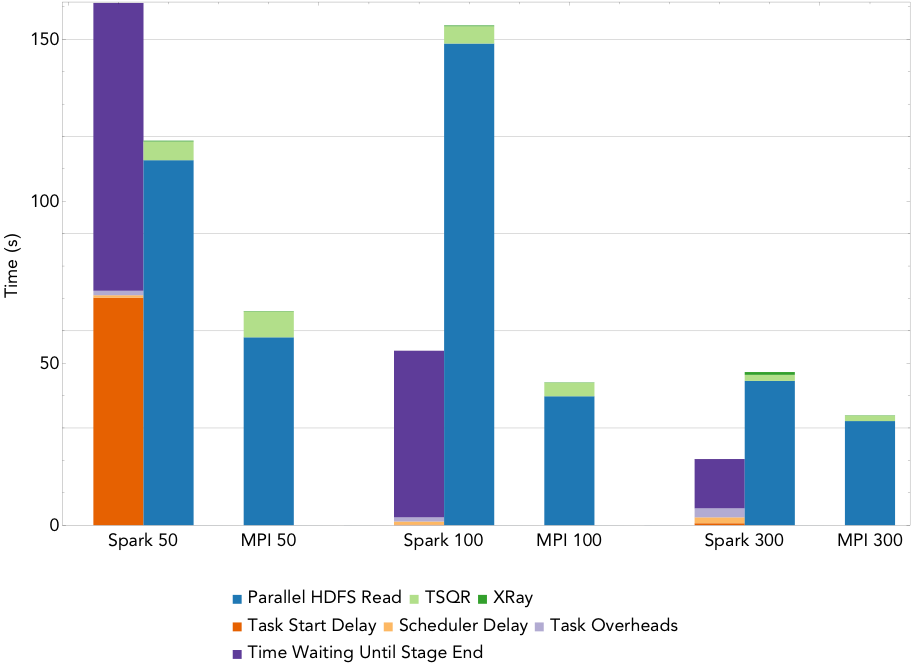}
\caption{Running time breakdown when using NMF to compute a rank 10 approximation to the 1.6TB Daya Bay matrix at 
  node counts of 50, 100, and 300. Each bin depicts the sum, over all stages, of the time spent in that bin by the average task within a stage. The 50 node run uses double the number of partitions as physical cores due to out-of-memory errors using fewer partitions-- this results in a large task start delay.}
\label{fig:nmfrt}
\end{center}
\end{figure*}

The TSQR algorithm used performs a single round of communication using a flat binary tree. Because there are few columns, the NMF algorithm is entirely I/O-bound. Figure~\ref{fig:nmfrt} gives the running time breakdown when computing rank 10 approximations using the MPI implementation of NMF on 50 nodes, 100 nodes, and 300 nodes. 
Each bin represents the sum, over all stages, of the time spent in that bin by the average task within a stage.

The running time for NMF is overwhelmingly dominated by reading the input. In comparison, TSQR and~\textsc{XRay} have negligible running times. Figure~\ref{fig:nmfrt} shows that the HDF5 read time does not scale linearly with the number of nodes and is the primary source of inefficiency -- this is due to saturating the system bandwidth for 72 OSTs. \textsc{XRay}, which is computed on the driver, is a sequential bottleneck and costs ~$100$ms at all node counts. TSQR only improves by tens of milliseconds, costing $501$ms, $419$ms, and $378$ms on 50, 100, and 300 nodes, respectively. This poor scaling can be attributed to hitting a communication bottleneck. Forming the TSQR binary tree is expensive for small matrices, especially using flat MPI. We did not tune our TSQR reduction tree shapes or consider other algorithms since TSQR is not the limiting factor to scalabilty. These results illustrate the importance of I/O scalability when performing terabyte-scale data parallel analytics on a high-performance architecture using MPI.

Figure~\ref{fig:nmfrt} also illustrates the running time breakdown for the Spark implementation of NMF on 50, 100, and 300 nodes. Unlike the MPI implementation, the Spark implementation incurs significant overheads due to task scheduling, task start delays, and idle time caused by Spark stragglers. For the 50 node run we configured Spark to use double the number of partitions as physical cores because we encountered out-of-memory errors using fewer partitions--- this incurs a task start delay overhead because some only half of the total tasks can be executed concurrently. The number of partitions was not doubled for the 100 and 300 node runs, so the task start delay overhead is much smaller for these runs. Similar to the MPI results, most of the running time is spent in I/O and Spark overheads, with a small amount of time spent in TSQR and \textsc{XRay}. Figure~\ref{fig:nmfrt} shows that the Spark implementation exhibits good strong scaling behavior up to 300 nodes.  Although the NMF algorithm used is entirely data parallel and suitable for Spark, we observed a $4\times$, $4.6\times$, and $2.3\times$ performance gap on 50, 100, and 300 nodes, respectively, between Spark and MPI. There is some disparity between the TSQR costs but this can be attributed to the lack of granularity in our Spark profiling, in particular the communication time due to Spark's lazy evaluation. Therefore, it is likely that the communication overhead is included in the other overhead costs whereas the MPI algorithm reports the combined communication and computation time.

\begin{figure}[h]
\centering
\includegraphics[width=.9\textwidth]{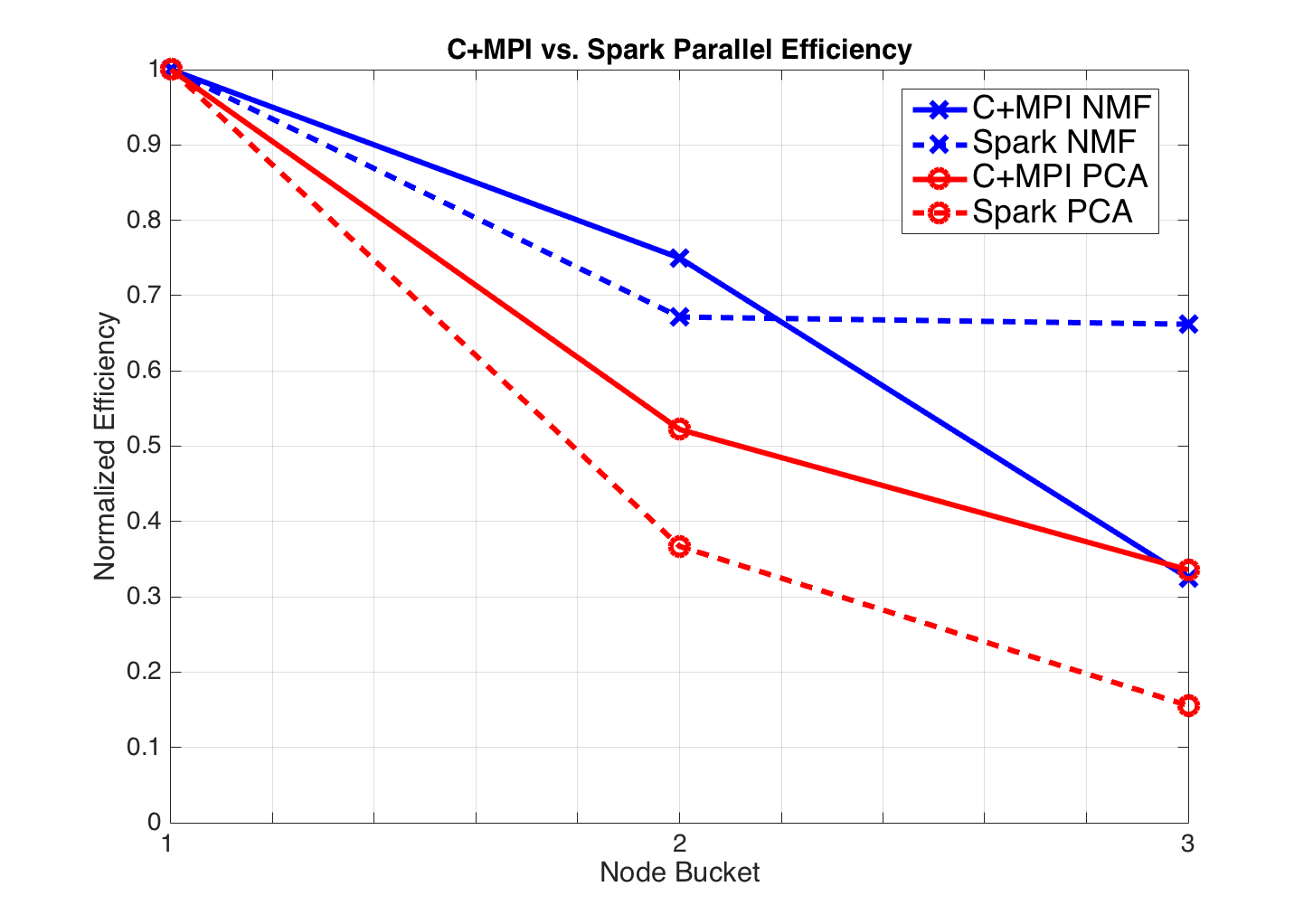}
\caption{Comparison of parallel efficiency for C+MPI and Spark. The x-axis label ``Node Bucket'' refers to the node counts. For NMF these are 50, 100, and 300 nodes (left to right) and 100, 300, and 500 nodes for PCA. For both algorithms, efficiency is measured relatively to the performance at the smallest node count.}
\label{fig:peff}
\end{figure}

Figure~\ref{fig:peff} shows the parallel efficiencies of the MPI and Spark implementations of NMF, normalized to the 50 node running time of the respective parallel frameworks. MPI NMF is completely dominated by I/O and the results are primarily indicative of scaling issues in the I/O subsystem. Spark NMF displays good scaling with more nodes; this is reflected in the parallel efficiency. However, the scaling is due primarily to decreases in the Spark overhead.

\paragraph{Science Implications.}
We are currently investigating the results of the NMF decomposition. Preliminary analysis indicates that we will need to augment the input data with non-linear features to make the input signals invariant to rotations and translations. Our eventual goal is to learn event-specific classifiers from the loadings of the NMF basis vectors. The classification will enable us to accomplish the final goal of segmenting and classifying the timeseries of sensor measurements. While implementing and verifying the scientific value of the entire pipeline is out of scope for this report, we have demonstrated the ability to apply our Spark NMF implementation to the TB-sized Daya Bay matrix. Together with feature augmentation, this will enable us to explore more advanced methods in the near future.

\subsection{PCA applied to the climate matrices}

We compute the PCA using an iterative algorithm whose main kernel is a distributed matrix-vector product. Since matrix-vector products are data parallel, this algorithm fits nicely into the Spark model. Because of the iterative nature of the algorithm, we cache the data matrix in memory to avoid I/O at each iteration.

\begin{figure*}[th!]
\centering
\includegraphics[width=.9\textwidth]{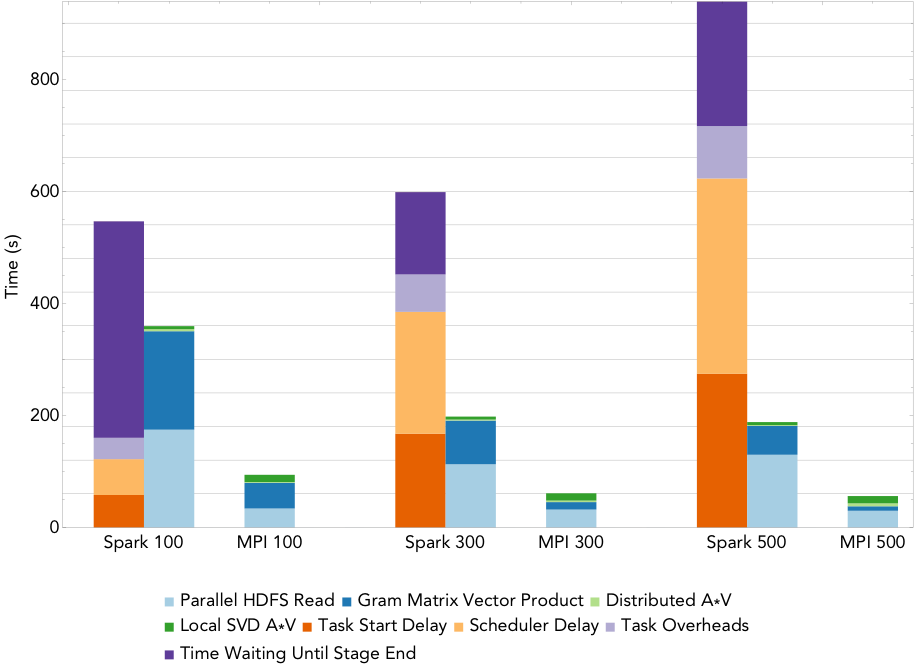}
\caption{Running time breakdown of PCA on the 2.2TB Ocean matrix at node counts of 100, 300 and 500. Each bin depicts the sum, over all stages, of the time spent in that bin by the average task within a stage.}
\label{fig:pcart}
\end{figure*}

\begin{figure}[th!]
\centering
\includegraphics[width=.9\textwidth]{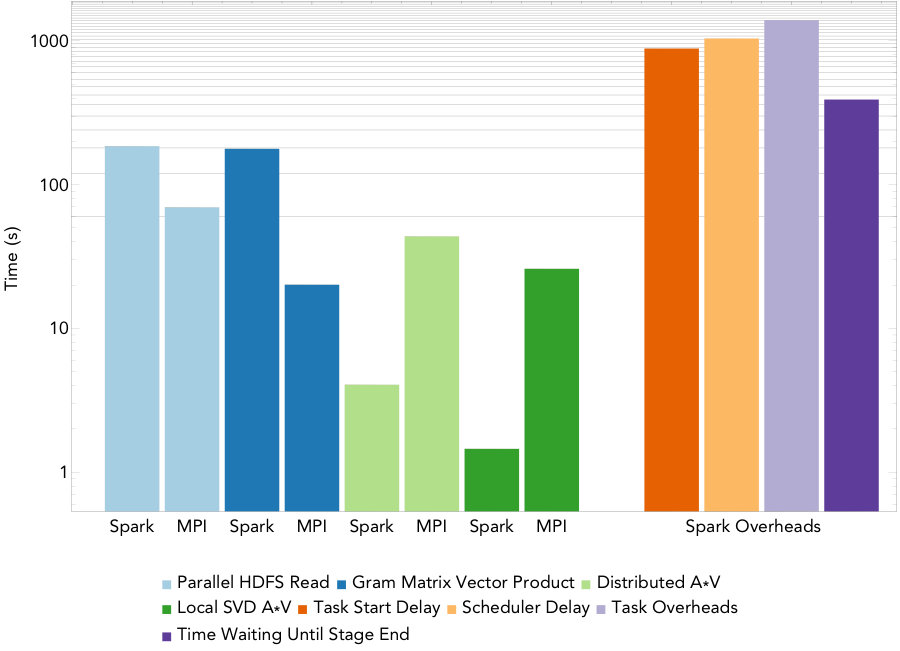}
\caption{Running time comparison of the Spark and MPI implementations of PCA on the 16TB Atmosphere matrix. Each bin depicts the sum, over all stage, of the time spent in that bin by the average task within a stage.}
\label{fig:hero}
\end{figure}

\paragraph{C+MPI vs. Spark.}
Figure~\ref{fig:pcart} shows the running time breakdown results for computing a rank-20 PCA decomposition of the Ocean matrix on 100, 300, and 500 nodes using the MPI implementation. Each bin depicts the sum, over all stages, of the time spent in that bin by the average task within a stage.

I/O is a significant bottleneck and does not exhibit the scaling observed for NMF in Figure~\ref{fig:nmfrt}. The I/O time is reduced going from 100 to 300 nodes, but not 300 to 500 nodes because the I/O bandwidth is saturated for the stripe size and number of OSTs used for the Daya Bay and Ocean data sets. The Gram matrix-vector products are a significant portion of the running time but scale linearly with the number of nodes. The matrix-matrix product ($AV$) does not scale due to a communication bottleneck. The bottleneck is because we compute a rank-$20$ PCA which makes communicating $V$ expensive. This cost grows with the number processors since it is entirely latency dominated. The final SVD of $AV$ is a sequential bottleneck and does not scale. Unlike NMF the sequential bottleneck in PCA is significant; future implementations should perform this step in parallel.

Figure~\ref{fig:pcart} also shows the scaling and running time breakdown of the Spark PCA implementation for 100, 300, and 500 nodes. The Gram matrix-vector products scale linearly with the number of nodes, however this is outweighed by inefficiencies in Spark. At this scale, Spark is dominated by bottlenecks due to scheduler delays, task overhead and straggler delay times. Task overhead consists of deserializing a task, serializing a result and writing and reading shuffle data. The Spark scheduling delay and task overhead times scale with the number of nodes, due to the centralized scheduler used in Spark. The iterative nature of the PCA algorithm stresses the Spark scheduler since many tasks are launched during each iteration. Under this workload we observed a 10.2$\times$, 14.5$\times$, and 22$\times$ performance gap on 100, 300, and 500 nodes, respectively, between Spark and MPI. The disparity between the costs of the $AV$ products and sequential SVDs in MPI and Spark can be attributed to the lack of granularity in our Spark profiling, in particular the communication time due to Spark's lazy evaluation. Therefore, it is likely that the communication overhead is included in the other overhead costs whereas the MPI algorithm reports the combined communication and computation time. 

Figure~\ref{fig:peff} shows the parallel efficiency of MPI PCA and Spark PCA. We observed that the MPI version hits an I/O bottleneck, a communication bottleneck in the $AV$ product and a sequential bottleneck in SVD($AV$). All of these are limiting factors and introduce inefficiencies to MPI PCA. Spark PCA is less efficient than MPI PCA due to scheduler delays, task overhead and straggler effects. The scheduler delays are more prominent in PCA than in NMF due to the larger number of tasks. NMF makes a single pass over the data whereas PCA makes many passes over the data and launches many tasks per iteration.

\paragraph{PCA Large-Scale Run.}
We used all 1600 Cori nodes to compute a rank-20 PCA decomposition of the 16TB Atmosphere matrix. In order to complete this computation in Spark in a reasonable amount of time, we fixed the number of iterations for the EVD of $A^TA$ to $70$ iterations. MPI PCA was able to complete this run in $160s$. Unfortunately we were unsuccessful at launching Spark on $1600$ nodes; after many attempts we reduced the number of nodes to $1522$. At this node count, Spark PCA successfully completed the run in $4175s$. Figure~\ref{fig:hero} shows the head-to-head running time comparison for this full-system run; each bin depicts the sum, over all stages, of the time spent within that bin by the average task within a stage. The Gram matrix-vector products are an order of magnitude more costly in Spark. We noticed that the tree-aggregates were very slow at full-system scale and are the likely cause of the slow Gram matrix-vector products. The $AV$ product and SVD are much faster in Spark than in MPI due to limited profiling granularity. Finally, we observed that the Spark overheads were an order of magnitude larger than the communication and computation time.

\paragraph{Science Implications.}
For the 2.2TB Ocean data set, the first two temporal ``empirical orthogonal functions'' (EOFS)--- corresponding to right singular vectors--- fully capture the annual cycles. The remaining time series show abrupt changes due to the 1983 El Ni\~no Southern Oscillation (ENSO), and more significantly, the record-breaking ENSO of 1997--98. The intermediate modes contain a complex interplay of various timescales, which is currently under investigation. The spatial EOFs, corresponding to the left singular vectors, show the relative dominance of the Indian Ocean Dipole and the classic warm pool--cold tongue patterns of ENSO at various depths below the ocean surface. Because of the 3D nature of the EOFs, we are able to see that the dynamic near the thermocline is most dominant, rather than that closer to the surface. Further, there are several smaller scale features that have a strong influence at different depths. Work is on-going to understand the nature of these different spatial patterns and the factors that influence their relative dominance.
 
\subsection{CX on the Mass-spec matrix}
\begin{table}[t]
\centering
\begin{tabular}{|c|c|c|c|} \hline
Algo & Size & \# Nodes & Spark Time (s)\\ \hline
\multirow{3}{*}{CX} & \multirow{3}{*}{1.1 TB} & $60$ & $1200$\\
{} & {} & $100$  & $784$\\
{} & {} & $300$ & $542$\\ \hline
\end{tabular}
\caption{Spark CX running times}
\label{tab:cxscale}
\end{table}
Much like PCA, the CX decomposition requires a parallel Gramian multiply, a distributed matrix-matrix product and a randomized SVD in order to compute extremal columns of $A$. CX is applied to the sparse 1.1TB MSI matrix, which is stored in the Parquet format. Table~\ref{tab:cxscale} shows the running times and scaling behavior of Spark CX. We found that Spark exhibited good scaling for the range of nodes tested and attained speedups of $1.5\times$ and $2.2\times$ on 100 and 300 nodes, respectively. The corresponding parallel efficiencies are 90\% for 100 nodes and 44\% for 300 nodes. These results show that the behavior of CX is similar to that of PCA, which is due to the overlap in their linear algebra kernels.

\paragraph{Science Interpretation.}
 The CX decomposition selected ions in three narrow regions of $m/z$. Among those ions identified as having significant leverage scores are ions at $m/z$ values of 439.0819, 423.0832, and 471.1276, which correspond to neutral losses of $\rm{CH_2}$, $\rm{CH_2O}$, and a neutral ``gain'' of $\rm{H_2O}$ from the 453.0983 ion.  These relationships indicate that this set of ions, all identified as having significant leverage scores, are chemically related.  That fact indicates that these ions may share a common biological origin, despite having distinct spatial distributions in the plant tissue.  

\subsection{Summary of Spark vs. C+MPI performance comparison}
We have demonstrated that matrix factorizations (which have traditionally been implemented using high-performance parallel libraries) can be implemented on Spark, and that Spark can run on large node counts on HPC platforms. By exploring the performance trade-offs of Spark matrix factorizations and comparing to traditional MPI implementations we have gained insights into the factors affecting Spark's scalability. Table~\ref{tab:matrix} summarizes the wall-clock times of the MPI and Spark implementations of the considered factorizations, and Table~\ref{tab:perfgaps} summarizes the performance gaps between Spark and MPI. These gaps range between $2\times - 25\times$ when I/O time is included in the comparison and $10\times - 40\times$ when I/O is not included. These gaps are large, but our experiments indicated that Spark I/O scaling is comparable to MPI I/O scaling, and that the computational time scales. The performance gaps are due primarily to scheduler delays, straggler effects, and task overhead times. If these bottlenecks can be alleviated, then Spark can close the performance gap and become a competitive, easy-to-use framework for data analytics on high-performance architectures.

\begin{table}[tbh]
\centering

\begin{tabular}{|c|c|c|c|c|c|} \hline
Algo & Size & \# Nodes & MPI Time (s) & Spark Time (s)\\ \hline
\multirow{3}{*}{NMF} & \multirow{3}{*}{1.6 TB} & $50$ & $66$ & $278$\\
{} & {} & $100$  & $45$ & $207$\\
{} & {} & $300$ & $30$ & $70$\\ \hline
\multirow{4}{*}{PCA} & \multirow{3}{*}{2.2 TB} & 100 & 94 & 934\\
 {} & {} & 300 & 60 & 827\\
 {} & {} & 500 & 56 & 1160\\ \cline{2-5}
 {} & {16 TB} & {MPI: 1600 Spark: 1522} & 160 & 4175 \\ \hline
\end{tabular}
\caption{{Summary of Spark and MPI running times.}}
\label{tab:matrix}
\end{table}

\begin{table}[tbh]
\begin{center}
\begin{tabular}{|c|c|c|c|} \hline
Algo & \# Nodes & Gap with I/O & Gap without I/O\\ \hline
\multirow{3}{*}{NMF} & $50$ & $4\times$ & $21.2 \times$\\
{} & $100$  & $4.6\times$ & $14.9\times$\\
{} & $300$ & $2.3\times$ & $15.7\times$\\ \hline
\multirow{4}{*}{PCA} & 100 & $10.2\times$ & $12.6\times$\\
 {} & 300 & $14.5\times$ & $24.7\times$\\
 {} & 500 & $22\times$ & $39.3\times$\\ \cline{2-4}
 {} & {MPI: 1600 Spark: 1522} & $26\times$ & $43.8\times$\\ \hline
\end{tabular}
\end{center}
\caption{Summary of the performance gap between the MPI and Spark implementations.}
\label{tab:perfgaps}
\end{table}

%% file: text/sparklessons.tex
Throughout the course of these experiments, we have learned a number of lessons pertaining to the behavior of Spark for linear algebra computations in large-scale HPC systems. 
In this section, we share some of these lessons and conjecture on likely causes.

\paragraph{Spark Scheduling Bottlenecks.}
The Spark driver creates and sends tasks serially, which can cause bottlenecks at high concurrency.  This effect can be quantified by looking at two metrics: Task Start Delay and Scheduler Delay. Task Start Delay measures the time from the start of stage until the task is sent to an executor. Scheduler Delay measures the additional time until the driver receives confirmation that the task has been received and its execution has started. Figure~\ref{fig:hero-timeline} is a plot of a sample of the tasks from one stage of the 16TB Spark PCA run. Note that the ordering of the colored bars within each task line does not correspond to the order they occurred---Spark uses a pipelined execution model, where different portions of a task are interleaved at a fine grain, and reports the total time spent on each activity.  We can see that the scheduling bottleneck causes a uniform distribution of start times, with tasks starting as late as 20 seconds after the earliest task.  The scheduler delay grows with the start delay, indicating that confirmation messages are queuing up and waiting to be processed at the driver when it finishes sending new tasks.

\begin{table}[th]
\centering
\begin{tabular}{| c | c | c | c | c | c | c |}
\hline
Algo & Size & Nodes & Partitions & Time (s) & Measured & Predicted \\
{} & {} & {} & {} & {} & Task Start & Delay \\
{} & {} & {} & {} & {} & Delay (s) & (2000/sec) \\
\hline
\multirow{4}{*}{PCA} & \multirow{3}{*}{2.2 TB} & 100 & 3200 & 924 & 411 & 112 \\
 {} & {} & 300 & 9600 & 827 & 332 & 336 \\
 {} & {} & 500 & 16000 & 1160 & 542 & 560 \\ \cline{2-7} & & & & & & \\[-1ex]
 {} & {16 TB} & 1522 & 51200 & 3718 & 1779 & 1792 \\
 \hline
\end{tabular}
\caption{Spark scheduling delays.}
\label{tab:scheduling}
\end{table}

Ousterhout et al.~\cite{Ousterhout13Sparrow} showed that these factors limit the Spark scheduler to launching approximately 1500 tasks per second.  Their measurements were based on an older version of Spark from 2013, but there have been no significant changes to the scheduler. Our results on Cori are consistent with a similar rate of about 2000 tasks per second.  We show the impact of this bottleneck on PCA in Table~\ref{tab:scheduling}.  
We expect that the largest negative impact on scaling is caused by the wait required to schedule the tasks in each iteration. The \emph{Measured Task Start Delay} column shows the sum of the largest task start delays in each Spark stage.  The \emph{Predicted Delay} column shows the delay predicted by a scheduling rate of 2000 tasks per second over 70 iterations and the listed number of tasks/partitions.  We observe that at 300, 500, and 1522 nodes, the task start delay is very close to the predicted~value.

This bottleneck represents a limit on the scaling achievable by Spark for highly iterative algorithms.  In particular, as the amount of parallelism increases, the minimum number of partitions and tasks also increases.  This results in a linearly increasing overhead from the scheduler as we increase parallelism.  This delay is further multiplied by the number of iterations.  We see the impact of this in the PCA results in Table~\ref{tab:scheduling}, where the final column represents this fixed overhead and is thus a lower bound on how fast we can execute at the given scale.  


\begin{figure}[tbhp]
\centering
\includegraphics[width=.6\textwidth]{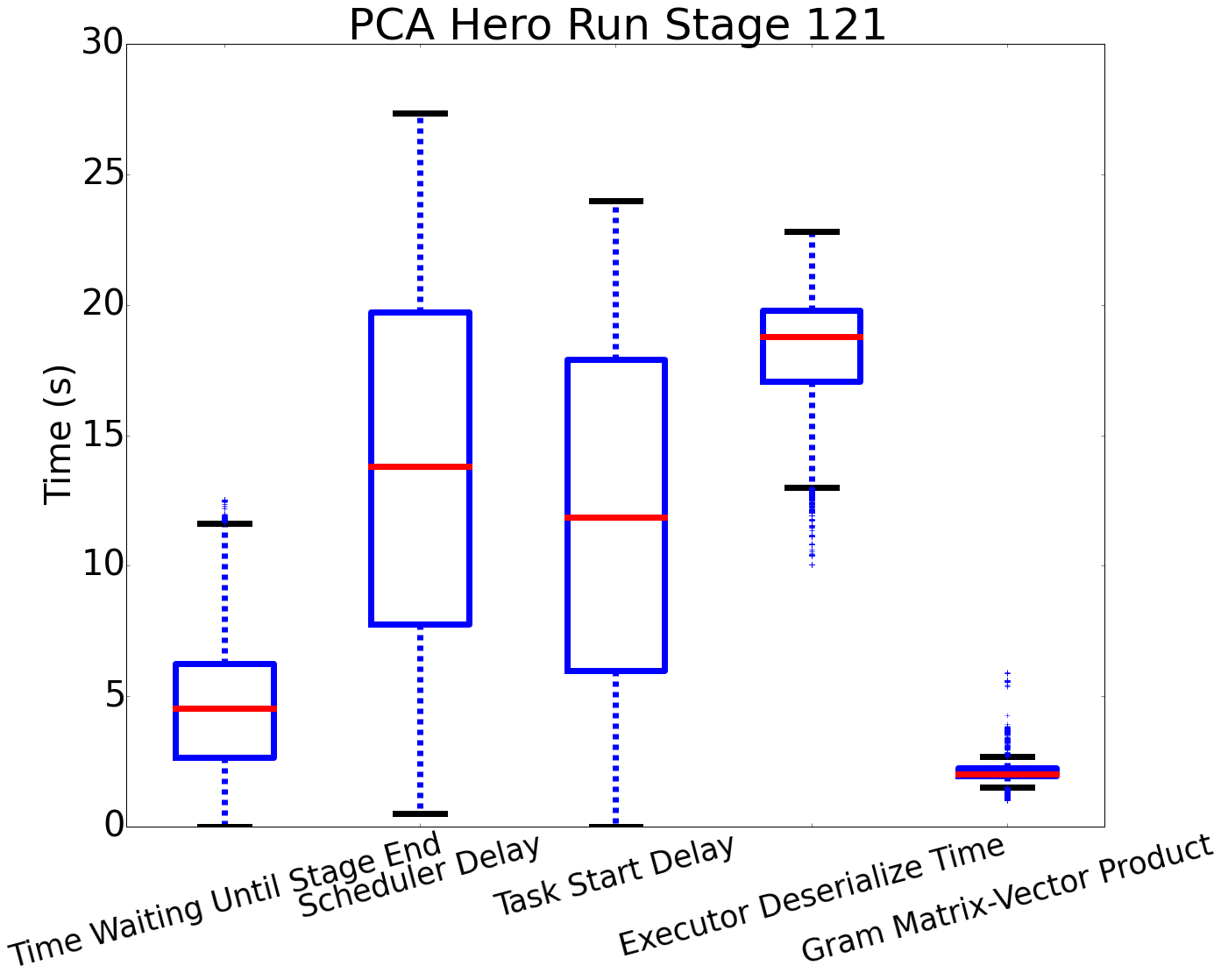}
\caption{Distribution of various components of all tasks in a multiply Gramian stage in the Spark PCA hero run. }
\label{fig:whisker}
\end{figure}

\paragraph{Other Significant Spark Overheads.}
Figures~\ref{fig:nmfrt} and~\ref{fig:pcart} illustrate that a large block of time is spent in Task Overheads. These overheads consist of the shuffle read and write time, the task deserialization time (executor deserialize time), and result serialization time.  During our runs on Cori, most of these overheads are insignificant with the exception of the executor deserialize time, as can be seen in Figure~\ref{fig:whisker}. High executor deserialize times are usually attributable to large tasks that take a long time to unpack. Also, any time spent in garbage collection during the deserialize phase counts toward the deserialize time.

\paragraph{Spark Variability and Stragglers.}
The time waiting for stage to end bucket in Figure ~\ref{fig:whisker} describes the idle time for a single stage in which a task has finished, but is waiting for other tasks to finish. The main cause of this idle time is what we call ``straggler effect", where some tasks take a longer than average time to finish and thus hold up the next stage from starting. In Figure~\ref{fig:whisker}, we can see there is some variability in the multiply Gramian component of the tasks, but this is insignificant compared to the remaining overheads. The straggler time may seem insignificant, however, Figure ~\ref{fig:whisker} shows the statistics for a single stage. When summed over all stages (i.e. all PCA iterations) the straggler effect does become significant overhead at O(100) seconds (see Figure \ref{fig:hero}).

The bulk-synchronous execution model of Spark creates scaling issues in the presence of stragglers. When a small number of tasks take much longer to complete, many cores waste cycles idling at synchronization barriers. At larger scales, we see increases in both the probability of at least one straggler, as well as the number of underutilized cores waiting at barriers.

During initial testing runs of the Spark PCA algorithm, variations in run time as large as 25\% were observed (in our staging runs we had a median run time of 645 seconds, a minimum run time of 489 seconds, and a maximum run time of 716 seconds). These variations could not be attributed to any particular spark stage. Sometimes the delay would occur in the multiply Gramian step, other times in the initial data collect stage. This variability is illustrated in the box and whiskers plot. Spark has a ``speculation" functionality which aims to mitigate this variability by restarting straggling tasks on a new executor. However, we found that enabling speculation had no appreciable effect on improving the run time, because the overhead to fetch a portion of the RDD from another worker was sufficiently high. This is because requests for RDDs from other workers must wait until the worker finishes its running tasks. This can often result in delays that are as long as the run time of the straggling task.

%% file: text/conclusion.tex
We conclude our study of matrix factorizations at scale with the following take-away messages: 
\begin{itemize}
  \item{A range of important matrix factorization algorithms can be implemented in Spark: we have successfully applied NMF, PCA and CX to TB-sized da tasets. We have scaled the codes on 50, 100, 300, 500, and 1600 XC40 nodes. To the best of our knowledge, these are some of the largest scale \emph{scientific data analytics} workloads attempted with Spark.}
\item{Spark and C+MPI head-to-head comparisons of these methods have revealed a number of opportunities for improving Spark performance. The current end-to-end performance gap for our workloads is $2\times - 25\times$; and $10\times - 40\times$ without I/O. At scale, Spark performance overheads associated with scheduling, stragglers, result serialization and task deserialization dominate the runtime by an order of magnitude.}
\item{{In order for Spark to leverage existing, high-performance linear algebra libraries, it may be worthwhile to investigate better mechanisms for integrating and interfacing with MPI-based runtimes with Spark. The cost associated with copying data between the runtimes may not be prohibitive.}}
\item{Finally, efficient, parallel I/O is critical for Data Analytics at scale. HPC system architectures will need to be balanced to support data-intensive workloads.}
\end{itemize}

%% file: text/ack.tex
This research used resources of the National Energy Research Scientific Computing Center, a DOE Office of Science User Facility supported by the Office of Science of the U.S. Department of Energy under Contract No. DE-AC02-05CH11231. 

We would like to thank Doug Jacobsen, Woo-Sun Yang, Tina Declerck and Rebecca Hartman-Baker for assistance with the large scale runs at NERSC. We thank Edgar Solomonik, Penporn Koanantakool and Evangelos Georganas for helpful comments and suggestions on tuning the MPI codes. We would like to acknowledge Craig Tull, Ben Bowen and Michael Wehner for providing the scientific data sets used in the study. 

This research is partially funded by DARPA Award Number HR0011-12-2-0016, the Center for Future Architecture Research, a member of STARnet, a Semiconductor Research Corporation program sponsored by MARCO and DARPA, and ASPIRE Lab industrial sponsors and affiliates Intel, Google, Hewlett-Packard, Huawei, LGE, NVIDIA, Oracle, and Samsung.  This work is supported by Cray, Inc., the Defense Advanced Research Projects Agency XDATA program and DOE Office of Science grants DOE DE-SC0010200 DE-SC0008700, DE-SC0008699. AD is supported by the National Science Foundation Graduate Research Fellowship under Grant No. DGE 1106400. Any opinions, findings, conclusions, or recommendations in this paper are solely those of the authors and does not necessarily reflect the position or the policy of the sponsors.